\documentclass[aps,prd,preprint,superscriptaddress]{revtex4}

\usepackage{latexsym}
\usepackage{graphicx}
\usepackage{epsfig}
\newcommand \be {\begin{equation}}
\newcommand \bea {\begin{eqnarray}}
\newcommand \ee {\end{equation}}
\newcommand \eea {\end{eqnarray}}
\newcommand \bed {\begin{displaymath}}
\newcommand \eed {\end{displaymath}}

\newcommand{\bit}{\begin{itemize}}
\newcommand{\eit}{\end{itemize}}

\begin{document}

\title{Energetic model of tumor growth }
\author{Paolo Castorina}
\author{Dario Zappal\`a}
\affiliation{INFN, Sezione di Catania,
\quad and \quad
Dept. of Physics, University of Catania,\\
Via S. Sofia 64, I-95123, Catania, Italy
}
\date{\today}
\begin{abstract}
A macroscopic model of the tumor Gompertzian growth is proposed.
This approach is based on the energetic balance among the different cell activities, 
described by methods of statistical mechanics 
and related to the growth inhibitor factors. The model is successfully applied 
to the multicellular tumor spheroid data.
\end{abstract}

\pacs{87.10.+e  87.17.-d  87.18.-h}

\maketitle

\section{Introduction}
A microscopic model of tumor growth in vivo is still an open problem.
However, in spite of the  large set of potential parameters
due to the variety of the in situ conditions, tumors have a peculiar growth pattern
that is generally described by a Gompertzian curve \cite{gompertz}, often considered 
as a pure phenomenological fit of the data.
More precisely there is an initial  exponential growth
(until 1-3 mm in diameter) followed by the vascular Gompertzian phase\cite{libro}.
Then it seems reasonable to think that cancer growth follows a general pattern that one can hope to describe
by macroscopic variables and following this line of research, 
for example, the universal model proposed in \cite{west} has been recently applied to
cancer\cite{torino}.
In this talk  we present a macroscopic model of tumor growth,
proposed in \cite{noi2}, that: 
i) gives an energetic basis to the Gompertzian law;
ii) clearly distinguishes between the general evolution patterns, 
which include the internal feedback effects, and the external constraints;
iii) can  give indications on the different tumor phases during its evolution.
The proposed macroscopic approach is not in competition with microscopic models \cite{varie}, but it 
is a complementary instrument for the description of the tumor growth.

\section{Cellular energetic balance and Gompertzian growth}
The Gompertzian curve is solution of the  equation 
\be
\label{gomp}
 \frac{dN}{dt}= N \gamma \ln\left( \frac{N_{max}}{N} \right)
\ee
where $N(t)$ is the cell number at time $t$, $\gamma$ is a constant and $N_{max}$ is the theoretical saturation value for
$t \to \infty$. 

It is quite natural to identify the right hand side of  Eq. (\ref{gomp})  as the  number of proliferating cells at time 
$t$ and then to consider $f_p(N)= \gamma \ln\left( \frac{N_{max}}{N} \right)$ as the fraction of proliferating cells
and  $1-f_p(N)=f_{np}$  the fraction of non proliferating cells.
Since $f_p(N)$ depends on $N(t)$, there is a feedback mechanism usually described  by introducing some 
growth inhibitor factors which increase with the number of non-proliferating cells  and 
are responsible for the saturation of the tumor size.
The concentration of
inhibitor factors should be  proportional to the number of non-proliferating cells which is maximum at $N=N_{max}$ \cite{suterland}.
If one considers that, during the growth, each  cell  shares out
its available energy at time $t$, in the average,  among its metabolic activities,
the mechanical work ( associated with the change of the tumor size and shape) and the increase of the number of cells, it 
is conceivable  to translate the previous cellular  feedback effect  in terms of energy content.
Indeed, as shown in \cite{noi2}, the specific energy for the growth should be proportional to $f_{np}$ 
and the average metabolic plus  mechanical  energy per cell, $M_e$, is proportional to $f_{p}$.
As we shall see this reproduces the observed cellular feedback. 

The model \cite{noi2}, based on an analogy with statistical mechanics, 
assumes that  in a larger system $B$, the body, there is a subsystem $A$, the tumor, 
made of $N(t)$ 
cells at  time $t$, with total energy   $U$, which  has specific distributive mechanisms for providing,
in the average, the amount of energy $U/N$ to each cell.
Then we indicate with  $E_M$  the energy needed for the metabolic activities of $A$, with $\Omega$ the energy associated 
with the mechanical work required for any change of size and shape  of $A$, and with $\mu$ the specific energy (i.e. per cell)
correlated to the change in the number of cells $N$, and, by assuming that these three processes summarize the whole 
cellular activity, we have $U=E_M + \Omega + \mu N$. 
Let us assume that the system $A$ slowly evolves through states of equilibrium with the system $B$, defined 
by macroscopic variables,  analogous for instance to the inverse temperature $\beta$, 
that have the same value for the two systems, 
although it should be clear that in our case we do not have  {\it real} thermodynamical equilibrium because  
the system $B$  supplies the global energy for the slow evolution of the subsystem $A$. 
Within  this scheme,
there are many microscopic 
states of the system $A$, compatible with the macroscopic state, 
defined by  $\beta$, $\mu$  
and  $V$,  which are built by a large number of states of each single cell.
These microscopic states of each cell have minimum total energy $\epsilon$ and
in an extremely simplified picture, we  assume an energy spectrum of the form $\epsilon_l = \epsilon  + l \delta$, where $l$ is an integer and
$\delta$ is the minimum energy gap between two  states. With this spectrum the grand partition function ${ Z}$
is given by the following product
${Z}(\beta,V,\mu)=\Pi_{l=0}^\infty {\rm exp}\left( e^{-\beta(\epsilon_l-\mu)} \right )$
and the corresponding grand potential, which is natural to associate to 
the energy $\Omega$ related to the mechanical work in our problem,  is given by
$\Omega(\beta,V,\mu)
= - (R/ \beta)  {\rm exp}[ {-\beta(\epsilon-\mu)}]$
where $R=1/(1-e^{-\beta\delta})$.
The average value of $N$, defined for constant $V$ and $\beta$, 
turns out to be $N=-\beta \Omega$.
According to  the basic rules of statistical mechanics, the product of the ``entropy times the temperature'',
which in our system corresponds to  $E_M$ introduced above, is
\be\label{eemme}
E_M=\left (\beta \frac{\partial \Omega}{\partial \beta}\right)_{V,\mu}=\frac{N}{\beta}
\left ( 1+C+\ln\left( \frac{R}{N} \right ) \right )
\ee
where $C$ is given by $C=R\beta\delta~{\rm exp}(-\beta\delta)$.
From the previous equations it is straightforward to express $\mu$ in terms of $N$ :
$\mu=\epsilon+(1/ \beta) \ln(N/R)$.
 
To find the evolution of the system $A$ which takes into account the internal feedback mechanism we 
recall  some results obtained in \cite{noi2} :

\begin{enumerate}
\item the energetic balance requires that the growth with cellular  feedback 
starts at a minimum number of cells  $N_m$ and saturates at 
a maximum value, $N_{max}$ related by $ N_{max} =N_m {\rm exp} (1 +\beta \epsilon)$.

\item  For $N_m >> 1$, 
$M_e= E_M/N+\Omega/N =(1/ \beta)  \ln (N_{max}/N)$ is a decreasing 
function of $N>N_m$ and
there is a simultaneous reduction of the total  metabolic energy per cell and
an increase of specific energy required for the growth:
there is an energetic balance between $M_e$
and $\mu=(1/\beta) ( 1 +  \beta \epsilon +\ln (N/N_{max}))$. Moreover $M_e$
is proportional to $f_p(t)$.
\end{enumerate}

Then it is possible to derive the Gompertz equation for the growth, $\Delta N$ in an interval $\Delta t$.  
$\Delta N$  is proportional to the number of proliferating cells and then one can write
$\Delta N=  c_1~\Delta t~ f_p(t) N =c_2 ~\Delta t~ M_e~N$
where $c_1$ and $c_2$ are constants. This gives in the continuum limit 
Eq. (\ref{gomp}) with $\gamma=c_2/\beta$.

\section{Application to Multicellular Tumor Spheroids}
The first step to analyze the phenomenological implications of the model
and to describe the dependence of the growth on the external conditions is to
consider the multicellular tumor spheroids (MTS). A minimal MTS description 
consists of a spherical growth where 
a) the thickness, $k$, of the layer where
the nutrient and oxygen is delivered (the crust) is independent on the  spheroid radius $R$;
b) the cell density is constant;
c) the cells in the crust, receive a constant supply of nutrient for cell;
d) at time t the cells are non proliferating if they are at  distances $d <R-k$ if $R>k$
from the center of the spheroid. For $R<k$ all cells are proliferating.

To separate the effects of the external constraints due to energy supply from those related to  biomechanical
conditions, it is better to consider first the MTS growth without external and internal  stress and to introduce 
later these effects.
\subsection{Energetic MTS growth}\label{sub}
In this case the external conditions
are experimentally  modified by changing the oxygen and nutrient concentration in the environment.
At fixed value of these concentrations, 
the maximum allowed number of  cells in the MTS is $N_{max}$.
For $R<k$ all  cells receive the nutrient and oxygen supply while for $R>k$ there is a fraction of 
non proliferating cells and the
feedback effect starts. The growth of the MTS  is due to the proliferating cells in the crust and 
one obtains that the MTS radius $R$, for $R >>k$, 
follows a Gompertzian law as the one in Eq. (\ref{gomp}), with $N$ replaced by $R$ and $ N_{max}$ by  $R_{max}$,
where ${R_{max}}$ is the maximum radius of the spheroid  corresponding to the maximum number of tumor cells $N_{max}$. 
The experimental results show that after 3-4 days of initial exponential growth the spheroids essentially follow the 
Gompertzian pattern\cite{suterland}. 

According to our model, at time $t >t^*$, such that $R(t^*)=k$, a fraction of the total cells becomes non proliferating,
the feedback effect starts and the growth rate decreases according to the Gompertz law.
The number of cells at time $t^*$ is fixed by the condition $N(t^*)=N_m$ \cite{noi2}.
On the other hand, the variation of the concentration of nutrient and/or of oxygen modifies the total energy supply, 
that is the value of $N_{max}$ and, since  $N_m= N_{max} {\rm exp}(-\epsilon \beta -2)$,
there is a clear correlation among the external energetic "boundary conditions", 
the value $N_{max}$ and the thickness of the viable cell rim which corresponds to the radius of the onset of necrosis.
It can be shown  \cite{noi2} that ($G_c$ is the glucose concentration):
\be\label{power}
k(G_c)=\alpha \left ( N_{max}^{1/3}- {N_{max}^{0}}^{1/3}\right )+ k_0
\ee
where $\alpha$ and $k_0$ are constants depending on the supplied oxygen. 
From Eq. (\ref{power}) one obtains the 
correlation among $N_{max},~G_c$ and $k$.
In Fig.1 and  Fig. 2 the previous behaviors are compared with data 
without optimization of the parameters.

\begin{figure}[hp] 
\centerline{\epsfxsize=8 cm \epsfbox{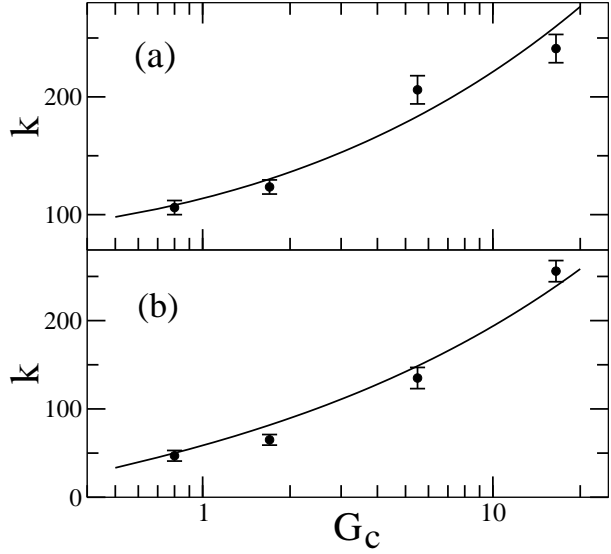}}   
\caption{Thickness ($\mu m$) vs. glucose concentration ($mM$).
Figure (a) is for an oxygen concentration of $0.28~mM$ 
and Figure (b) is for an oxygen concentration of $0.07~mM$.
 } 
\end{figure}

\begin{figure}[hp] 
\centerline{\epsfxsize=8 cm\epsfbox{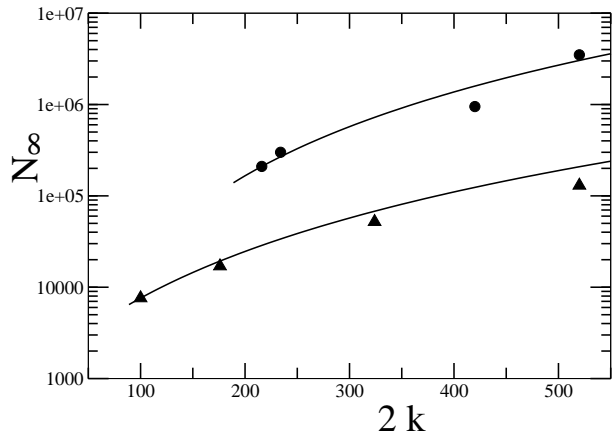}}   
\caption{Spheroid saturation cells number vs. diameter ($\mu m$) at which necrosis first develops.
Circles  refer to culture in $0.28~mM$ of oxygen 
Triangles  refer to culture in  $0.07~mM$  of oxygen. } 
\end{figure}

\subsection{ Biomechanical effects}
The experimental data indicate that  when  MTS are under a solid stress, obtained for example  by a gel, 
the cellular density $ \rho$ is not constant and depends on the external gel concentration $C_g$. 
In particular the results in \cite{hel}
show that: 1) an increase of the gel concentration 
inhibits the growth of MTS;
2) the cellular density at saturation increases with the gel concentration.
In the model  the mechanical energy is included in the energetic balance of the system by the term 
 $\Omega =  - N/\beta = - PV$  where   the  pressure is $P(t)= \rho(t)/\beta$.
The introduction of this term decreases
the value of  $N_{max}$ with respect to the case  in Sect.~\ref{sub} and
this reduction should also imply
a decrease of the maximum size of the spheroids, i.e. $ R_{max}(P)$
by increasing the pressure.
The comparison with the data is reported in Table I for $C_g$ in the range $0.3 - 0.8~\%$
(see \cite{noi2} for details).

\begin{table}[bt]
{\footnotesize
\begin{tabular}{@{}crrrr@{}}
\hline
{} &{} &{}\\[-1.5ex]
$C_g$ (percent)& $2R_{max}(\Omega)$ $[\mu m]$  exper.& $2R_{max}(\Omega)$ $[\mu m]$ fit \\[1ex]
\hline
{} &{} &{}\\[-1.5 ex]
0.3  &  450  &  452 \\[1ex]
0.5  &  414  &   429\\[1ex]
0.7  &  370  &   404\\[1ex]
0.8  &  363  &  394\\[1ex]
\hline
\end{tabular}\label{table1}}
\caption{\rm
Comparison with the experimental data as discussed in the text. The experimental error is about
$\pm 10 \%$.}
\end{table}


\begin{thebibliography}{0}

\bibitem{gompertz} B. Gompertz, {\it Phyl. Trans. R. Soc.} , {\bf 115}, 513 (1825).

\bibitem{libro} G.G. Steel, ``Growth Kinetic of tumors'', Oxford Clarendon Press, 1977;
 {\it Cell tissue  Kinet.},  {\bf 13}, 451 (1980);
T.E. Weldon, ``Mathematical models in cancer research'', Adam Hilger Publisher, 1988
and refs. therein.
\bibitem{west} G.B. West {\it et al.},  {\it Nature}  {\bf 413}, 628  (2001).
\bibitem{torino} C. Guiot {\it et al.},  {\it J. Theor. Biol.} {\bf 25}, 147  (2003).

\bibitem{noi2} P. Castorina and D. Zappal\`a, ``Tumor Gompertzian growth by cellular 
energetic balance'', q-bio.CB/0407018.

\bibitem{varie} M. Marusic  {\it et al.},  {\it Cell Prolif.} {\bf 27} 73  (1994); 
Z. Bajzer  {\it et al.}  in:
``Survey of model for tumor-immune system dynamics'', 
J.A. Adams and N. Bellomo eds.,  Birkhauser 1997;
A. Bru  {\it et al.},  {\it Phys. Rev. Lett.}  {\bf 81}, 4008 (1998);
Z. Bajzer,  {\it Growth Dev. Aging} , {\bf 63}, 3 (1999);
N. Bellomo {\it et al.},  ``Mathematical topics on the modelling complex mulicellular
systems and tumor immune cells competition'',  Preprint: Politecnico di Torino, 2004.
\bibitem{suterland} J. P. Freyer, R.M. Sutherland,  {\it Cancer Research},  {\bf 46},
3504 (1986).
\bibitem{param} L. Norton {\it et al.},  {\it Nature} {\bf 264}, 542 (1976);  
L. Norton  {\it Cancer Research} ,  {\bf 48}, 7067 (1988).
\bibitem{hel} G. Helmlinger  {\it et al.},  {\it Nature Biotechnology} ,{\bf 15},  778 (1997).
\end{thebibliography}
\end{document}